\documentclass[10pt,aps,prd,nofootinbib,superscriptaddress, twocolumn,preprintnumbers,balancelastpage]{revtex4}
\usepackage{amssymb,amsmath,latexsym,graphics, graphicx,epsfig,multirow,comment,hyperref,appendix}

\newcommand{\gsim}{\lower.7ex\hbox{$\;\stackrel{\textstyle>}{\sim}\;$}}
\newcommand{\lsim}{\lower.7ex\hbox{$\;\stackrel{\textstyle<}{\sim}\;$}}

\newcommand{\Msun}{\text{ M}_{\odot}}
\newcommand{\zre}{\text{z}_{\text{re}}}

\usepackage[usenames,dvipsnames]{color}
\definecolor{orange}{RGB}{220,120,0}

\begin{document}

\title{Dark Matter Debris Flows in the Milky Way}

\author{Mariangela Lisanti}
\affiliation{PCTS, Princeton University, Princeton, NJ 08540}

\author{David N. Spergel}
\affiliation{Department of Astrophysical Sciences, Princeton University, Princeton, NJ 08540}

\begin{abstract}
We show that subhalos falling into the Milky Way create a flow of tidally-stripped debris particles near the Galactic center with characteristic speed behavior, but no spatial features.  Using the Via Lactea-II N-body simulation, we study the unvirialized component arising from particles that were bound in subhalos around the time of reionization but have since been tidally stripped.  These debris particles constitute a few percent of the local density today and have speeds peaked at $340$ km/s in the solar neighborhood.  This spatially-homogeneous velocity substructure has important implications for surveys of low-metallicity stars, as well as direct detection experiments sensitive to dark matter with large scattering thresholds.
\end{abstract}
 \maketitle
 
In the $\Lambda$CDM paradigm, structure formation in the early universe begins roughly twenty million years after the Big Bang with the gravitational collapse of matter around density fluctuations \cite{Peebles:1982ff}.  The first virialized objects are Earth-mass halos, which continue to grow in size, reaching $10^6$--$10^8 \Msun$ around $z=10$ \cite{Diemand:2005vz}.  Dark matter halos of this size host active star formation at their cores.  These early stars are the pioneers of future galaxy growth for the largest halos, but become sparse relics in the less massive halos, which experience a dampening in star formation rates as gas is photoevaporated during reionization \cite{Bullock:2000qf, Bullock:2000wn, Kravtsov:2004cm}.  In contrast to their larger counterparts, these small halos remain dark.

The Milky Way halo forms hierarchically by accreting these small, selfbound subunits  \cite{Searle:1978gc}.  While some of these halos merge completely and virialize, others do not and continue to orbit the Galaxy as satellites.  These subhalos experience tidal forces leading to substantial mass loss, particularly on pericentric passage when the forces are most extreme~\cite{Hayashi:2003sj,Diemand:2007qr}.  The stripped particles become a part of the Galactic halo and those with shared progenitors sometimes retain distinctive phase-space features.  The stars that formed near the cores of the subhalos may also be stripped, leading to spatial or velocity features in the stellar halo~\cite{Helmi:2008eq, Johnston:1996sb, Johnston:1997fv, Johnston:2012yh}.   

Substructure in star distributions can provide hints for potential structure in the dark matter distribution, as has already been the case for stellar streams associated with the disruption of dwarf galaxies.  Stellar streams are spatially aligned and have small velocity dispersions~\cite{Stiff:2001dq,Johnston:1996sb, Johnston:1997fv, Helmi:1999ks, Helmi:2002iu, Bullock:2005pi}.  Their presence strongly suggests that dark matter streams also formed from material stripped from infalling satellites.  A dark matter stream is dynamically cold, has a one-dimensional morphology, and can enhance the modulation fraction of a signal at a direct detection experiment~\cite{Kuhlen:2009vh,Freese:2003tt, Gelmini:2004gm}.  Studies of dark matter streams in numerical simulations, however, suggest that a single dark matter stream that dominates the local density is rare~\cite{Vogelsberger:2009bn, Vogelsberger:2010gd, Maciejewski:2010gz}.       

In this Letter, we introduce another class of velocity substructure that is \emph{spatially homogeneous}.  
While spatially-homogeneous velocity substructure has been discussed with regards to the stellar halo (see~\cite{Helmi:2008eq} for a review), it has not received any attention with regards to dark matter.  However, it can provide a powerful probe for experiments because a generic velocity feature that is nearly spherically-distributed will be guaranteed in the solar neighborhood.  Using the Via Lactea-II (VL2) N-body simulation, we will show that a subcomponent of the dark matter near Earth is comprised of tidal debris with a distinct high-speed behavior, but no characteristic spatial features. 
We refer to this velocity substructure as a ``debris flow,'' to distinguish it from a tidal stream.

Via Lactea-II is a high-resolution N-body simulation that follows the evolution of a billion $4.1\times10^3 \Msun$ particles from $z=104$ to the present \cite{Diemand:2008in, Diemand:2007qr, Zemp:2008gw}.  It assumes a $\Lambda$CDM universe with the best-fit cosmological parameters from the WMAP-3 year data release~\cite{Spergel:2006hy}.  The simulation is centered on an isolated halo with no recent major merger that might be a suitable candidate for a Milky Way-like galaxy; the host has a mass of $1.94\times10^{12} \Msun$ and a tidal radius of 462 kpc at $z=0$.  There are 20047 subhalos identified at $z=0$ in the VL2 simulation with peak circular velocity larger than 4 km/s.  Of these, nearly all have progenitors at $\zre = 9$~\cite{VLwebsite}, which we choose as the characteristic time when baryon cooling in the subhalos becomes inefficient due to the onset of reionization.  We study tidal debris from this subset of subhalos because of its relevance for substructure in the old stellar halo; a more complete study of debris from the full catalogue of VL2 subhalos is explored in follow-up work~\cite{Lisanti:tocome}.  

\begin{figure}[b] 
   \centering
   \includegraphics[width=3.5in]{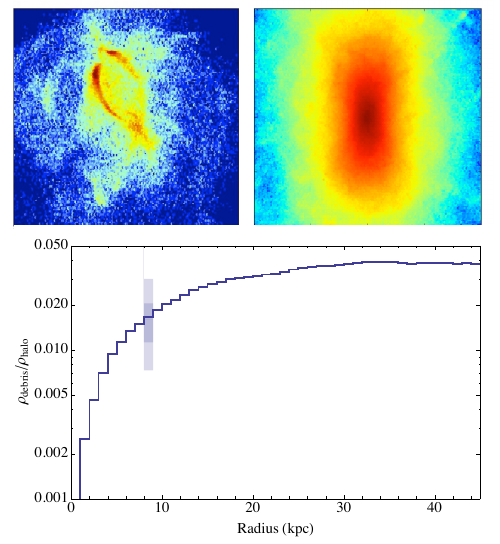} 
   \caption{(Top left) Distribution of dark matter particles at present time that were stripped from an orbiting satellite with a mass of $5.6\times10^7\Msun$ at $\zre=9$.  (Top right) The distribution at present time of all tidally-stripped particles.  Both distributions are projections in the x-z plane, centered on the Galactic origin and extending out a radial distance of 100 kpc.  (Bottom) Fractional density of the debris particles relative to the Milky Way halo as a function of radius.  The spread in density in the 8--9 kpc bin is indicated by the shaded bars (see text).} 
   \label{fig: debris}
\end{figure}

We identify all particles bound to the subhalo progenitors at $\zre$ and then determine their corresponding positions and velocities in the present epoch.  A particle is labeled as ``debris'' if it is not bound to a subhalo today.  
The top left panel of Fig.~\ref{fig: debris} is a snapshot at $z=0$ of the debris from a randomly chosen orbiting satellite.  The tidally-stripped particles are distributed within a $100$ kpc radius of the Galactic center.  There are relatively larger densities of particles near the turning points of the orbits.  Notice, in particular, that the density is very high within $\sim10$ kpc of the center, where particles are stripped on the satellite's pericentric passage.  In this region, individual streams cannot be distinguished and the overdensity is fairly uniform.  The top right panel of Fig.~\ref{fig: debris} shows the particles that have been tidally-stripped from \emph{all} the infalling satellites.  The debris shows no spatially-localized structure, but rather follows the mass-contours of the prolate VL2 halo~\cite{Diemand:2009bm}.

The subhalos at $\zre$ range in mass from $10^5 - 10^9 \Msun$, and roughly 4000 of these progenitors (approximately 20\%) contribute debris particles today within 45 kpc of the Galactic center.  Of these, the most massive halos contribute the largest fraction.  In particular, about half of the debris consists of particles from halos with mass greater than $5\times10^7\Msun$, while $20\%$ comes from halos with mass greater than $4\times 10^8\Msun$ at $\zre$.  

\begin{figure}[tb] 
   \centering
   \includegraphics[width=3.5in]{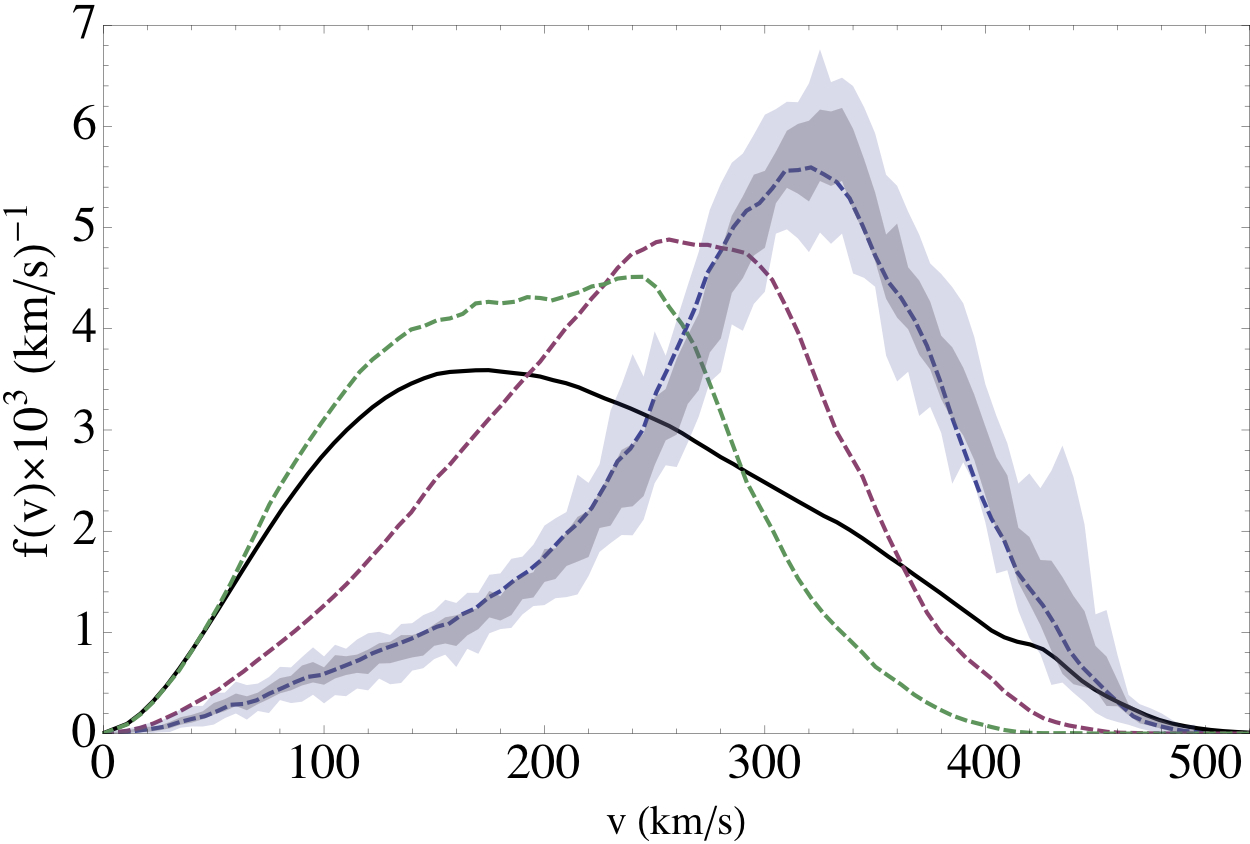} 
   \caption{Speed distribution of the dark matter debris.  The black line is the distribution of all VL2 particles in a shell from $r = 5$--$15$  kpc.   The dotted lines are the distributions for debris particles located 30--45 (green), 15--30 (pink), and 5--15 (blue) kpc from the Galactic center.  The dark blue region is the $1\sigma$ spread about the mean for 100 sample spheres with radius 5 kpc centered at $r=10$ kpc.  The light blue region shows the maximum and minimum for the same sample spheres. }
   \label{fig: veldist}
\end{figure}

Figure~\ref{fig: veldist} shows the speed distributions for debris particles located 30--45 (green), 15-30 (pink), and 5--15 (blue) kpc from the Galactic center.  For comparison, the solid black curve shows the distribution for all VL2 particles in a 5--15 kpc shell.  Note that these distributions are separately normalized to emphasize the differences in their shapes.  The speed of the debris particles increases for radial shells closer to the Galactic center.  In the 5--15 kpc shell, the speeds of the debris particles are peaked at $\sim340$ km/s and the shape of the debris distribution is distinct from that of all the VL2 particles in this shell.  

The radial dependence of the debris speeds is a simple consequence of energy conservation.  The gravitational potential is related to the density distribution of the halo, which, assuming an NFW-like profile~\cite{Navarro:2008kc,Navarro:1996gj,Zhao:1995cp}, is the following for the VL2 parent halo:
\begin{equation}
\rho(r) = \frac{\rho_s}{(r/r_s)^{\gamma}(1+r/r_s)^{3-\gamma}},
\end{equation}
where $\rho_s = 3.5\times10^{-3}\Msun\text{pc}^{-3}$, $r_s = 28.1$ kpc, and $\gamma = 1.24$ are the best-fit parameters for the local density, scale radius, and inner radius, respectively~\cite{Diemand:2009bm}.  The initial velocity of the debris is taken to be the average velocity of the parent subhalos at final apocenter.  Using these assumptions for the initial conditions and the potential, we estimate that $v \approx 275$ km/s at $r = 25$ kpc and $v \approx 370$ km/s at $r = 10$ kpc; these speeds correspond remarkably well with Fig.~\ref{fig: veldist}. 

\begin{figure*}[t] 
   \centering
   \includegraphics[width=7in]{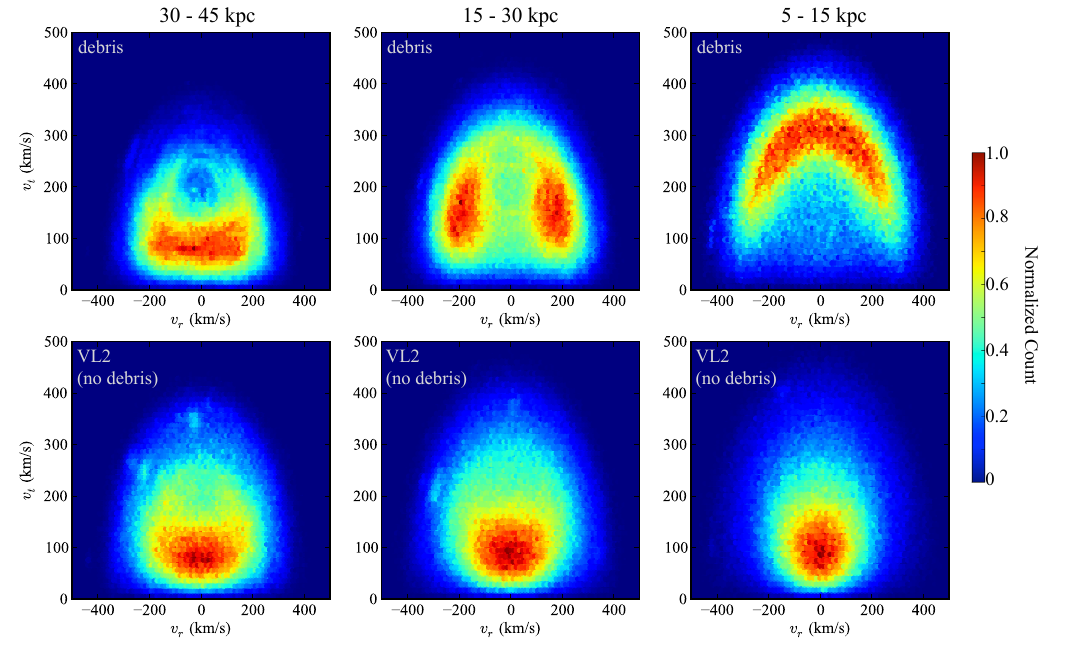} 
   \caption{Tangential vs. radial velocity for debris particles within $30 < r < 45$ kpc (left), $15 < r < 30$ kpc (middle), and $5< r < 15$ kpc (right) of the Galactic center.  The top row shows the distributions for the debris particles, while the bottom row shows the distributions for all other particles in VL2.}
   \label{fig: vrvtdist}
\end{figure*}

It is clear, then, that the speed behavior of the debris can be explained by the common origin of the particles in infalling subhalos.  The fact that these particles are tidally stripped from satellites also affects the radial, $v_r$, and tangential, $v_t$, velocity distributions.  The top row of Fig.~\ref{fig: vrvtdist} shows the regions of highest density in the $v_r - v_t$ plane for the debris particles within 30--45 kpc (left), 15--30 kpc (middle) and 5--15 kpc (right).  There is a dramatic increase in the tangential velocity of the debris particles from $\sim 60$ km/s far from the Galactic center up to $\sim$ 340 km/s within the inner 15 kpc of the halo.  At large radii, the radial velocity is centered at $v_r = 0$ and has small dispersion.  It is maximized between 20 and 30 kpc, with a value of $\sim$ 180 km/s.  In the inner 15 kpc, $v_r$ is again centered at 0, but with a much larger dispersion.  The corresponding distributions for all non-debris particles in these radial shells are shown in the bottom row for comparison.    

The velocities of the debris particles closest to the Galactic center have a large tangential component.  This is because these particles are tidally-stripped at or near the pericenter of their parent halo's orbit at small radii.  Further from the Galactic center, the velocities of the debris flow have a larger radial component because the particles are stripped from infalling satellites that have not yet reached their turning points.  A random selection of late-falling high-speed particles in the Milky Way should not necessarily exhibit this radial dependence in $v_r$ and $v_t$.  Indeed, the behavior observed in Fig.~\ref{fig: vrvtdist} is a direct consequence of the tidal origin of the debris.

To determine whether the dark matter debris has important observational consequences, it is necessary to characterize both its density and velocity behavior in the solar neighborhood.  The bottom panel of Fig.~\ref{fig: debris} shows the density of the debris relative to the VL2 halo, as a function of radius.  From 5--20 kpc, where the speed behavior is most distinct from the rest of the halo, the debris comprises 1-3\% of the Milky Way.  This fraction increases to 4\% at 40 kpc; however, as shown in Fig.~\ref{fig: vrvtdist}, the speeds of the debris flow are not as high at these distances.  From 7.5--9.5 kpc, the debris makes up about 1.7\% of the halo.  The fractional contribution of debris in this radial shell increases to about 4.4\% for particles with Earth-frame speeds greater than 500 km/s (in June).

To determine the local variation in the debris flow density, we find the relative density in a hundred sample spheres with radius 0.5 kpc centered at 8.5 kpc.  The dark blue bar in Fig.~\ref{fig: debris} shows the 1$\sigma$ spread and the light blue extends over the total range of sampled densities.  Part of this variation is due to the fact that the samples are taken in a spherical shell, but the VL2 halo is prolate~\cite{Diemand:2009bm}.    

To characterize local deviations in the debris' speed behavior, we sample the speed distributions for a hundred spheres centered at r = 10 kpc with 5 kpc radius.  The dark blue region in Fig.~\ref{fig: veldist} is the $1\sigma$ spread in the distributions over these samples, and the light blue region shows the minimum and maximum value found in each speed bin.  The spread in the debris' speed distribution has some localized peaks, but its overall shape remains remarkably consistent over the entire spherical shell.  

In summary, a subcomponent of the local Milky Way halo is characterized by dark matter tidal debris with unique speed behavior, but no local spatial structure.  We introduce the term ``debris flow'' for this class of spatially-homogeneous velocity substructure.  Debris flows
\begin{itemize}
\item consist of overlapping sheets, streams, plumes and shells created by dark matter tidally stripped from infalling subhalos.
\item have a spatial distribution indistinguishable from the background halo.
\item have a peaked speed distribution and unique radial and tangential velocity behavior, as determined by the orbital properties of the subhalo progenitors. 
\end{itemize}  
The debris in VL2 that was originally bound in subhalos at the time of reionization now constitutes approximately a few percent of the local density, and has speeds peaked $\sim 340$ km/s in the solar neighborhood.  

The debris flows described in this Letter serve as one of the first examples of spatially-homogeneous velocity substructure in the Milky Way halo, and should be studied in different simulations.  For example, the fractional density of the debris depends on the number of subhalos resolved by VL2 and will vary between simulations with different resolutions and initial conditions.  Despite these variations, the debris flow should be generic to similar high-resolution Milky Way simulations like GHalo~\cite{Stadel:2008pn} and Aquarius~\cite{Springel:2008cc, Maciejewski:2010gz}, whose subhalo concentrations are in good agreement with VL2 \cite{Diemand:2009bm}.  The properties of the debris flow will also be affected by the inclusion of the Galactic disc.  VL2 does not include baryonic physics, which will increase the internal velocities of the orbiting satellites, as well as the energies of the debris particles.

It is important to remember the selection bias in the subhalos considered in this work.  In particular, the particles labeled as ``debris'' originate from subhalos that still exist in the present epoch and are bound to the Milky Way.  The debris does not include particles stripped from subhalos that either pass through the Milky Way or are completely destroyed during infall.  In addition,  we do not consider particles that were bound at redshifts other than $z_{\text{re}}=9$.  Debris from this redshift provides a good starting point for understanding debris flows because of its relevance for star surveys, as well as direct detection experiments.  However, contributions from other redshifts should increase the relative density of dark matter debris and is explored in follow-up work~\cite{Lisanti:tocome}.  

The presence of debris flows can be experimentally verified in several different ways.  One possibility is to look for an imprint of the  flow in the local stellar distribution.  Because the time required for momenta exchange between stars is much longer than the age of the Galaxy, the kinematics of old stars encode information about their origin.  Stars that are tidally-stripped from subhalos ``trace out'' the paths of their dark matter counterparts~\cite{Helmi:2008eq}.  These stars are some of the oldest and most metal-poor in the Milky Way because they originate from subhalos with inefficient star formation after reionization.  Simulations of dark matter and baryonic evolution have followed the accretion of satellites in simple galactic models and have found evidence for a rich morphology of structure in the stellar halo~\cite{Johnston:2008fh}.  The specific evidence for debris flow would be metal-poor stars that exhibit distinct velocity behavior, but no distinct spatial features, over large areas of the sky.   

There is accumulating experimental evidence for  substructure from surveys that study both the photometric and spectroscopic properties of stars over large fields of view. 
One of the most dramatic examples is the discovery of the Sagittarius dwarf galaxy \cite{Ibata:1994fv} and the trail of stars that form a stream along its orbit \cite{Ivezic:2000ua, Yanny:2000ty, Ibata:2000pu, Ibata:2001iw, Johnston:1995vd}.  Several other spatial overdensities have been located in the stellar halo  \cite{Newberg:2001sx, Belokurov:2006kc, Grillmair:2006nx} and efforts have been made to detect structure in velocity-space~\cite{Majewski:1996zz, Helmi:1999uj, Schlaufman:2009jv}.  For example, a recent SEGUE study identified new elements of cold substructure from excesses in line-of-sight velocities over stellar halo expectations~\cite{Schlaufman:2009jv}.  The upcoming GAIA \cite{Perryman:2001sp} satellite will reconstruct the proper motions of stars and will be an integral step in mapping the velocity substructure in the solar neighborhood. 

Direct detection experiments, which look for nuclei recoiling from a collision with a dark matter particle, provide an alternate test for debris flows.  In particular, the debris results in high-speed flows in the Earth frame that yield a distinctive energy spectrum for the recoiling nucleus~\cite{Lisanti:tocome}.  Models in which the dark matter has a minimum scattering threshold near the velocities of the debris particles would be particularly sensitive to this feature in the velocity distribution.  Light elastic dark matter \cite{Bottino:2003cz, Fornengo:2010mk,Petriello:2008jj, Kaplan:2009ag, Fitzpatrick:2010em, Chang:2010yk} is one example that has recently received considerable attention in the literature.

Debris flows offer a unique way to search for Galactic dark matter, with star surveys and direct detection experiments providing orthogonal detection possibilities.  A discovery would indicate that a significant fraction of the local halo is unvirialized and retains distinctive phase-space features, despite being spatially uniform.  More importantly, it would open a window to the long history of our  Galaxy's evolution, providing a fossil record of the subhalos that merged to form the Milky Way.  

\section*{Acknowledgements}
We are particularly grateful to Juerg Diemand for providing the Via Lactea raw data, and to Neil Weiner for his helpful feedback over the course of this work.  In addition, we thank Daniele Alves, Juerg Diemand, Michael Kuhlen, Nikhil Padmanabhan, Constance Rockosi, and Jay Wacker for useful discussions.  ML acknowledges support from the Simons Postdoctoral Fellows Program as well as the US National Science Foundation, grant NSF-PHY-0705682.

\bibliography{streamsBibFile}

\end{document}